\documentclass[twocolumn,floatfix,showpacs,superscriptaddress]{revtex4}
\usepackage{graphicx}
\usepackage{amsmath}
\usepackage{amssymb}
\usepackage{bm}

\begin{document}

\title{Metallic phase of the quantum Hall effect in four-dimensional space}
\author{J. M. Edge}
\affiliation{Instituut-Lorentz, Universiteit Leiden, P.O. Box 9506, 2300 RA Leiden, The Netherlands}
\author{J. Tworzyd{\l}o}
\affiliation{Institute of Theoretical Physics, Faculty of Physics, University of Warsaw, Ho\.{z}a 69, 00--681 Warsaw, Poland}
\author{C. W. J. Beenakker}
\affiliation{Instituut-Lorentz, Universiteit Leiden, P.O. Box 9506, 2300 RA Leiden, The Netherlands}
\date{June 2012}
\begin{abstract}
We study the phase diagram of the quantum Hall effect in four-dimensional (4D) space. Unlike in 2D, in 4D there exists a metallic as well as an insulating phase, depending on the disorder strength. The critical exponent $\nu\approx 1.2$ of the diverging localization length at the quantum Hall insulator-to-metal transition differs from the semiclassical value $\nu=1$ of 4D Anderson transitions in the presence of time-reversal symmetry. Our numerical analysis is based on a mapping of the 4D Hamiltonian onto a 1D dynamical system, providing a route towards the experimental realization of the 4D quantum Hall effect.
\end{abstract}
\maketitle

To understand our three-dimensional world it is helpful and rewarding to ask how physics would differ in other dimensions. The interest in two-dimensional and four-dimensional worlds goes back to the 19th century \cite{Rod73,Abb84,Hin84}, inspired by the development of non-Euclidean geometry. In modern times, nanotechnology has brought us the 2D ``Flatland'' in semiconductor heterostructures and in graphene \cite{Gei07}, while ultracold atomic lattices offer a way to simulate 4D dynamics by encoding the extra dimension in an internal degree of freedom of the atoms \cite{Boa12}. 

These studies of physics in other dimensions are particularly interesting if they reveal effects that lack a 3D counterpart. The integer quantum Hall effect is a celebrated example, first discovered in 2D \cite{Kli80}, then generalized to 4D and 8D \cite{Zha01,Ber03}, and now known to exist in any \textit{even}-dimensional space \cite{Men03,Qi08,Ryu10}. In each case an integer-valued topological invariant (a Chern number ${\cal C}$ \cite{Tho82,Qi08,Ryu10}) counts the number of gapless surface modes, observable as the quantized Hall conductance in 2D. Disorder localizes single-particle excitations in the bulk but not on the surface, producing an insulating quantum Hall fluid bounded by a conducting surface.

Topological phase transitions (at which ${\cal C}$ switches from one integer to another) have been extensively studied in the 2D quantum Hall effect \cite{Huc95}, but not for the higher dimensional generalizations. As we will show here, in 4D the quantum Hall insulators with a different topological invariant are separated in phase space by a metallic region, so that a topological phase transition is composed of a pair of metal-insulator transitions. In 2D an intermediate metallic phase requires the presence of time-reversal symmetry (as in the quantum \textit{spin}-Hall effect \cite{Ryu10b}), but in 4D there is no such requirement. We find that the 4D quantum Hall insulator-to-metal transition has a different critical behavior than the 4D Anderson transitions in time-reversally-symmetric systems \cite{Gar07,Gar08}.

To determine the phase diagram of the 4D quantum Hall effect we proceed as follows: We start from a Dirac Hamiltonian \cite{Zha01,Qi08} of electrons with 4D momentum $\bm{p}$ coupled to SU(2) spin and valley degrees of freedom $\bm{\sigma}$, $\bm{\tau}$. This model has time-reversal symmetry, so it is in the universality class of the quantum spin-Hall effect (class AII in the classification of Anderson transitions \cite{Eve08}). In order to bring it in the universality class A of the quantum Hall effect, we add time-reversal-symmetry breaking disorder. Following the lines set out in Ref.\ \cite{Dah11} for the 2D quantum Hall effect, we then map the 4D Hamiltonian problem onto a 1D dynamical system. Here we use this dimensional reduction to arrive at an efficient numerical simulation, but in principle this method could be used to simulate the 4D quantum Hall effect in ultracold atomic lattices \cite{Cha08}.

The Dirac Hamiltonian has the form \cite{Qi08,note1}
\begin{align}
&H(\bm{p}) =\frac{2\arctan|\bm{v}|}{|\bm{v}|}{}\sum_{i=0}^4 v_i\Gamma_i,\label{Hpdefa}\\
&\bm{v}=K\left(\mu+\sum_{i=1}^4 \cos p_i,\sin p_1,\sin p_2,\sin p_3,\sin p_4\right),\label{Hpdefb}\\
&\bm{\Gamma}=(\tau_x\otimes\sigma_0,\tau_z\otimes\sigma_0,\tau_y\otimes\sigma_x,\tau_y\otimes\sigma_y,\tau_y\otimes\sigma_z).\label{Hpdefc}
\end{align}
The vectors $\bm{v},\bm{\Gamma}$ have components $v_{i},\Gamma_{i}$ ($i=0,1,2,3,4$). The Pauli matrices $\sigma_{\alpha}$, $\tau_{\alpha}$ ($\alpha=x,y,z$), with unit matrices $\sigma_{0},\tau_{0}$, act on the spin and valley degrees of freedom. The two parameters $K,\mu$ are real numbers.

The Hamiltonian \eqref{Hpdefa} is a $4\times 4$ matrix dependent on the momentum vector $\bm{p}=(p_1,p_2,p_3,p_4)$, in the Brillouin zone $|p_{i}|<\pi$ of a 4D square lattice (lattice constant and $\hbar$ set equal to unity). It satisfies the time-reversal symmetry relation
\begin{equation}
H(\bm{p})={\cal T}H(-\bm{p}){\cal T}^{-1},\;\;{\cal T}=(i\sigma_{y}\otimes\tau_{x}){\cal K},\label{TRSdef}
\end{equation}
with ${\cal K}$ the operator of complex conjugation. The spectrum consists of a pair of twofold degenerate bands
\begin{equation}
E_{\pm}(\bm{p})=\pm 2\arctan|\bm{v}|,\label{Epmdef}
\end{equation}
with a band gap centered at the Fermi level $E=0$. The topological invariant (the second Chern number) depends on the effective mass parameter $\mu$, according to \cite{Qi08,Gol93}
\begin{align}
{\cal C}=
\begin{cases}
3\,{\rm sign}\,(\mu) &{\rm if}\;\; 0<|\mu|<2,\\
-\,{\rm sign}\,(\mu) &{\rm if}\;\; 2<|\mu|<4,\\
0 &{\rm if}\;\; |\mu|>4.
\end{cases}\label{calCdef}
\end{align}
Changes in ${\cal C}$ are signaled by a closing and reopening of the band gap at time-reversally invariant momenta [$p_{i}=-p_{i}$ (mod $2\pi$)], where $v(\bm{p})$ vanishes.

We now include a spin-dependent potential $V(\bm{x})$, with position operator $\bm{x}=(x_1,x_2,x_3,x_4)=i\partial/\partial\bm{p}$, to break both translational invariance and time-reversal symmetry. To arrive at a dynamical system the potential is added stroboscopically in time,
\begin{equation}
{\cal H}(t)=V(\bm{x})+H(\bm{p})\sum_{n=-\infty}^{\infty}\delta(t-n).\label{Htdef}
\end{equation}
The time evolution of a wave function over one period is given by the Floquet operator
\begin{equation}
{\cal F}=e^{-iH(\bm{p})}e^{-iV(\bm{x})}.\label{floquet_as_prod}
\end{equation}
(We measure position in units of the lattice constant and set the stroboscopic period equal to unity.) Periodically time-dependent models of this type are called quantum kicked rotators \cite{Izr90}. 

A first advantage of the stroboscopic description is that a simple separable potential,
\begin{equation}
V(\bm{x})=V_{1}(x_{1})-\sum_{i=2}^4 \omega_i x_i,\;\;V_{1}(x_{1})=x_{1}^{2}\,\tau_{z}\otimes\sigma_{0},\label{Vseparable}
\end{equation}
can be used to simulate $d$-dimensional disorder --- provided the frequencies $\omega_{i}/2\pi$ are incommensurate irrational numbers \cite{Bor97,note2}. Note that the $\tau_{z}$ matrix anticommutes with ${\cal T}$ from Eq.\ \eqref{TRSdef}, so this potential breaks time-reversal symmetry. 

A second advantage is that the $d$-dimensional dynamics with a single stroboscopic period can be mapped exactly onto a 1-dimensional dynamics with $d$ stroboscopic periods \cite{Cas89}. For that purpose we choose an initial wave function of the form
\begin{equation}
\Psi(\bm{p},t=0)=\psi(p_1,t=0)\prod_{i=2}^4\delta(p_i-\alpha_i).\label{Psit0}
\end{equation}
During one period the momentum $p_i$ ($i=2,3,4$) is incremented to $p_i+\omega_i$ (mod $2\pi$), so we may replace the 4D dynamics by a 1D dynamics with a time-dependent Floquet operator
\begin{equation}
{\cal F}(t)=e^{-iH(p_1,\omega_2 t+\alpha_2,\omega_3 t+\alpha_3,\omega_4 t+\alpha_4)}e^{-iV_{1}(x_{1})}.\label{Ftdef}
\end{equation}

We introduce a Bloch number $q$ and calculate the time dependence of the wave function $\psi(p_1,t) = e^{−iqp_1}\chi(p_1,t)$. The state $\chi(p_1,t)$ is a $2\pi$-periodic function of $p_1$, so it is a superposition of a discrete set of eigenstates $e^{−imp_1}$ of $x_1$. We truncate the set to $M$ states, $m\in \{1, 2, \ldots M\}$, with periodic boundary conditions at the end points. The calculation of $\psi$ at integer $t$ then amounts to subsequent multiplications of the Floquet operator, which can be done efficiently using the Fast Fourier Transform algorithm \cite{Dah11}. Ranges of $M$ up to $10^3$ and $t$ up to $6\cdot 10^7$ are included in the simulations.

\begin{figure}[tb] 
\centerline{\includegraphics[width=1.0\linewidth]{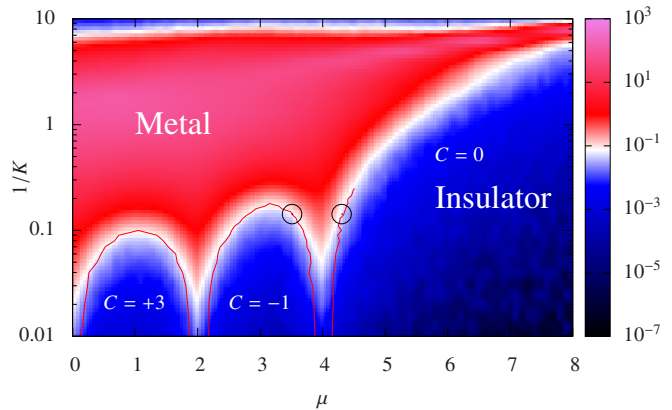}}
\caption{Phase diagram of the 4D quantum Hall effect (broken time-reversal symmetry, class A). The colour scale represents ${\cal D}=\Delta^2/\sqrt t$ at $t=10^5$. The phase boundary separating metal and insulator appears approximately at ${\cal D}\approx 0.08$ (white). A more precise determination of the phase boundary from the scaling analysis is indicated by the red curves (with circles marking the two transitions analyzed in Table \ref{tab:crit_expts}). Because of the $\pm\mu$ symmetry only positive values of $\mu$ are shown.
}
\label{fig:phase_diag_shepels_freqs}
\end{figure}

The metallic and insulating regions in the phase diagram are obtained by initializing $\chi(x_1,t)$ at $x_1=m_0$ and calculating the mean square displacement at time $t$,
\begin{equation}
\Delta^2(t)= \overline{\sum_m (m- m_0)^2 |\chi(x_1=m,t)|^2}.\label{def_of_delta}
\end{equation}
The overbar indicates an average over initial conditions and over the Bloch number $q$. In the large-$t$ limit one has $\Delta(t)\to {\rm constant}$ in the insulating phase (localization) and $\Delta^2(t)\propto t $ in the metallic phase (regular diffusion). On the critical line separating these two phases $\Delta^2(t)\propto \sqrt t $ (anomalous diffusion). The resulting phase diagram as a function of effective mass $\mu $ and disorder strength $1/K $ is shown in Fig.\ \ref{fig:phase_diag_shepels_freqs}.

The clean limit of the model is reached for $K\to\infty$ [because then $V(\bm{x})$ is negligibly small relative to $H(\bm{p})$]. In this limit the system switches between topologically distinct insulating phases at $|\mu|=0,2,4$, see Eq.\ \eqref{calCdef}. Disorder introduces a metallic phase in between the insulating phases, so that now a change in ${\cal C}$ happens via two metal-insulator transitions. At fixed $\mu$ the metallic phase exists for a finite range of disorder strengths, vanishing both for weak and strong disorder. The insulating phase that appears at the largest disorder strengths ($1/K\gtrsim 10$) is disconnected from the insulating phases at smaller disorder, but presumably it is topologically trivial (${\cal C}=0$).

\begin{figure}[tb] 
\centerline{\includegraphics[width=0.9\linewidth]{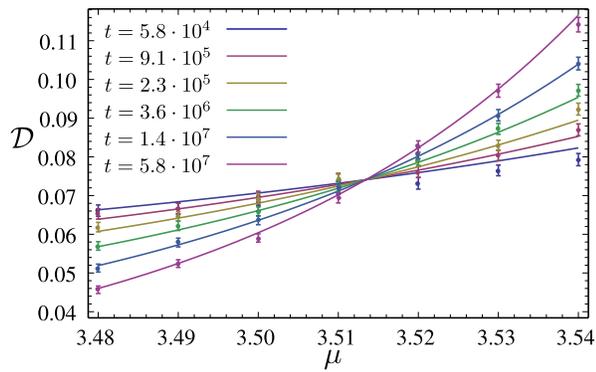}}
\caption{Finite-time scaling of the mean-square displacement at the quantum Hall insulator-to-metal transition (left circle in Fig.\ \ref{fig:phase_diag_shepels_freqs}). The data points result from the numerical simulation, the curves are fits to the scaling law \eqref{scalinglaw}. The crossing identifies the critical point $\mu_c$, separating the metallic phase ($\mu>\mu_c$) from the insulating phase ($\mu<\mu_c$). 
}
\label{fig:scaling}
\end{figure}

To characterize the metal-insulator transition we have performed a finite-time scaling analysis, along the lines of Refs.\ \cite{Dah11,Lem09}. One-parameter-scaling in 4D requires that the mean-square-displacement obeys the scaling law
\begin{equation}
\frac{\Delta^2(t)}{\sqrt t} = {\cal D}(\xi^{-4}t),\;\;\xi\propto |\mu-\mu_{c}|^{-\nu}.\label{scalinglaw}
\end{equation}
All microscopic parameters enter only through the localization length $\xi$, which has a power law divergence at the critical point $\mu_{c}$. We determine the critical exponent $\nu$ by expanding ${\cal D}$ in a power series near $\mu_c$, including also finite-time corrections to scaling \cite{Sle09}. A typical scaling plot is shown in Fig.\ \ref{fig:scaling} \cite{note4}.

\begin{table}[tb]
\begin{tabular}{ccccc}
symmetry & topological  & critical \\
class &  invariant ${\cal C}$ &exponent $\nu$ \\
\hline\hline
A &  $-1$	    &$1.18\pm 0.05$\\
A &  $0$	   &$1.21\pm 0.03$\\ 
\hline
AII  & $-1$	  &$0.99\pm 0.02$\\ 
 AII & $0$	 &$1.02\pm 0.04$\\ 
 \end{tabular}
\caption{Critical exponent $\nu$ for the metal-insulator transitions indicated by circles in Figs.\ \ref{fig:phase_diag_shepels_freqs} and \ref{fig:phase_diag_shepels_freqs-class_A2}. The value ${\cal C}$ of the topological invariant at the insulating side of the transition is indicated, as well as the symmetry class: class A (quantum Hall effect) or class AII (quantum spin-Hall effect).}
\label{tab:crit_expts}
\end{table}

Results are listed in Table \ref{tab:crit_expts} (first two rows). As expected from general considerations \cite{Ryu10b}, there is no dependence of $\nu$ on the value of the topological invariant ${\cal C}$ at the insulating side of the transition. Within numerical accuracy, $\nu\approx 1.2$ for the transition from a metal to either a trivial or a nontrivial quantum Hall insulator.

\begin{figure}[tb] 
\centerline{\includegraphics[width=1.0\linewidth]{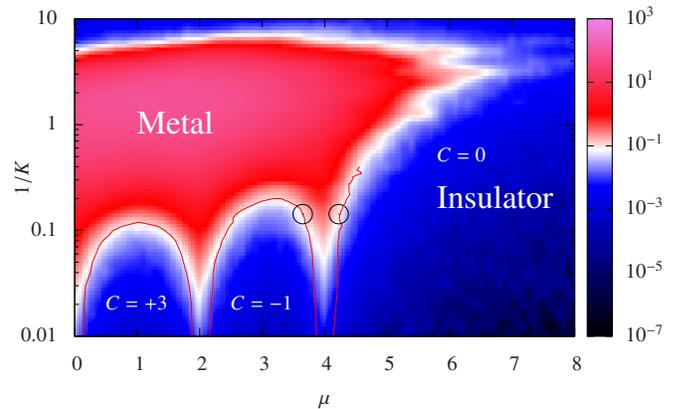}}
\caption{Phase diagram of the 4D quantum spin-Hall effect (preserved time-reversal symmetry, class AII). The colour scale is as in Fig.\ \ref{fig:phase_diag_shepels_freqs}.
}
\label{fig:phase_diag_shepels_freqs-class_A2}
\end{figure}

The calculations described so far are for a system with broken time-reversal symmetry, in class A of the quantum Hall effect. To investigate the role played by this symmetry we have repeated the calculations in the time-reversally-symmetric class AII of the quantum spin-Hall effect, by replacing the $\tau_z$ matrix in Eq.\ \eqref{Vseparable} with the unit matrix $\tau_{0}$. The potential $V(\bm{x})$ then commutes with ${\cal T}$ and preserves time-reversal symmetry. The phase diagram is qualitatively the same, see Fig.\ \ref{fig:phase_diag_shepels_freqs-class_A2}, at least for not too strong disorder. The insulating phase for $1/K\gtrsim 10$ is now connected to the ${\cal C}=0$ phase at weaker disorder, so we definitely know that this is a trivial insulator. 

Table \ref{tab:crit_expts} (last two rows) shows that a significant effect of time-reversal symmetry appears in the critical exponent $\nu\approx 1.0$ for class AII --- well below what we found for class A. Within numerical accuracy, the critical exponent which we find for the 4D quantum spin-Hall effect agrees with the semiclassical formula \cite{Gar07,Gar08,note3}
\begin{equation}
\nu_{\rm sc}=\frac{1}{2}+\frac{1}{d-2}\label{nusc}
\end{equation}
for $d$-dimensional Anderson transitions with time-reversal symmetry.

In conclusion, we have demonstrated that the phase diagram of the quantum Hall effect depends sensitively on the dimensionality. While in 2D only insulating phases exist, in 4D a metallic phase appears as well, for an intermediate range of not too strong and not too weak disorder. Topological phase transitions in the 4D quantum Hall effect occur via consecutive insulator-to-metal-to-insulator transitions, rather than via a single insulator-to-insulator transition as in 2D. This altogether different phenomenology is made possible by the fact that broken time-reversal symmetry prevents the formation of a metal in 2D but not in higher dimensions.

One direction for further theoretical research is towards a semiclassical scaling theory of 4D quantum Hall transitions. In 2D there is no hope for such a theory, but in 4D the situation is more promising in view of the close agreement between numerics and semiclassics that we have found for the critical exponent in the quantum spin-Hall effect. From the experimental point of view, the mapping of a scalar 3D Hamiltonian onto a 1D dynamical system has been realized in cold atoms with a pulsed optical lattice, driven by three incommensurate frequencies \cite{Cha08,Lem09}. Our 4D-to-1D mapping adds the complexity of an additional driving frequency, but more importantly would require control over an internal degree of freedom of the atoms to produce the required $4\times 4$ matrix structure.  

We thank A. R. Akhmerov, J. P. Dahlhaus, and E. P. L. van Nieuwenburg for useful discussions.
This research was supported by the Dutch Science Foundation NWO/FOM, by an ERC Advanced Investigator Grant, and by the EU network NanoCTM.

\appendix

\section{Finite-time scaling analysis of the metal-insulator transitions}
\label{scalingApp}

Here we give the details of the scaling analysis presented in the main text, following the established procedure \cite{Sle09}. Close to the transition point $\mu_c $, the anomalous diffusion coefficient ${\cal D}=\Delta^{2}/\sqrt{t}$ obeys the single-parameter scaling law \eqref{scalinglaw}, which we can write in the form
\begin{equation}
\ln{\cal D}=F(t^{1/4\nu}u),\;\;u=\mu-\mu_{c}+{\cal O}(\mu-\mu_c)^{2}.\label{calDF}
\end{equation}
Finite-time corrections to scaling add a term $t^{-y}G(t^{1/4\nu}u)$ to $\ln{\cal D}$, with $y>0$ the leading irrelevant exponent.

We expand the functions $F$ and $G$ in a power series in $t^{1/4\nu}u$, and expand $u$ in a power series in $\mu-\mu_c$,
\begin{align}
\ln{\cal D}&= \ln { \cal D}_c + \sum_{k=1}^{N_1} c_k^{(1)} (t^{1/4\nu} u)^k +\sum_{k=1}^{N_3} c_k^{(3)} t^{-y} (t^{1/4\nu} u)^{k-1}, \label{eq:1a}\\
 u&= \mu-\mu_c + \sum_{k=2}^{N_2} c_k^{(2)} (\mu-\mu_c)^k.
    \label{eq:1b}
\end{align}
We fit the unknown parameters to the numerical data for $\Delta^{2}(\mu,t)$, averaged over $500-1000$ initial conditions, varying $\mu$ around $\mu_c$ for six values of $t$ in the range $10^4-6\cdot 10^7$. The number of fit parameters $N_1,N_2,N_3 $ was increased until we reached a value for chi-squared per degree of freedom ($\chi^2$/ndf) close to unity. 

We have performed this scaling analysis for two sets of incommensurate frequencies, in order to make sure that our results for the critical exponents do not depend on the precise choice of driving frequencies. The two sets are\\
$\bullet$ Set $\Omega_1$:
  $\omega_2=2\pi/\lambda,\;\omega_3=2\pi/\lambda^2,\; \omega_4=2\pi/\sqrt 2$, where $\lambda$ is the real solution to the equation $x^3 -x-1=0$ \cite{Bor97}.\\
$\bullet$ Set $\Omega_2$:  
  $\omega_2=2\pi\ln3,\;\omega_3=2\pi\ln4,\; \omega_4=2\pi\ln5$.\\
Results for both frequency sets are given in Tables~\ref{tab:crit_expts_class_A} and \ref{tab:crit_expts_class_A2}, and summarized for frequency set $\Omega_1$ in Table \ref{tab:crit_expts}.

\begin{table}[tb]
    \begin{tabular}{cccccccccc}
     class& ${\cal C}$& freq.\ set&$\mu_c$&  $N_1$&$N_2$ &$N_3$		       &$\nu$ & $\chi^2$/ndf \\ \hline\hline
  A&          $-1$	 & $\Omega_1$ &$3.51$      &1&2&0&$1.18\pm 0.05$ & 0.97\\
  A&    $-1$	 & $\Omega_2$    &$3.51$      &2&1&0	       &$1.21\pm0.06$ & 1.01\\
  A&    0	 & $\Omega_1$   &$4.30$      &1&1&1        &$1.21\pm 0.03$ & 0.89\\
   A&   0	 & $\Omega_2$	 &$4.30$    &1&2&1       &$1.20\pm0.03$ & 1.05
    \end{tabular}
    \caption{Parameters for the scaling analysis of the metal-insulator transition in the 4D quantum Hall effect (symmetry class A).}
    \label{tab:crit_expts_class_A}
  \end{table}
  
\begin{table}[tb]
  \begin{tabular}{ccccccccccc}
    class&${\cal C}$& freq.\ set& $\mu_c$ & $N_1$&$N_2$ &$N_3$		       &$\nu$ & $\chi^2$/ndf \\ \hline\hline
   AII &     $-1$	 & $\Omega_1$    &$3.65$       &1&1&1	       &$0.99\pm0.02$ & 1.16\\
 AII &   $-1$	 & $\Omega_2$ &$3.62$   &1&1&1&$0.98\pm 0.04$ & 0.78\\
  AII &  0	 & $\Omega_1$  &$4.22$     &1&1&1        &$1.02\pm 0.04$ & 0.88\\
 AII &   0	 & $\Omega_2$ 	 &$4.24$      &1&2&2       &$1.07\pm0.05$ & 0.91
  \end{tabular}
  \caption{Parameters for the scaling analysis of the metal-insulator transition in the 4D quantum spin-Hall effect (symmetry class AII).}
  \label{tab:crit_expts_class_A2}
\end{table}


\begin{thebibliography}{99}
\bibitem{note1} The factor $(2/v)\arctan v$ in the Hamiltonian \eqref{Hpdefa} can be omitted without changing the topological invariant. It is included following Ref.\ \cite{Dah11}, to flatten the band structure and reduce finite-size effects.
\bibitem{note2} We performed our simulations for two sets of irrational incommensurate frequencies: $\{\omega_{2}/2\pi,\omega_{3}/2\pi,\omega_{4}/2\pi\}$ = $\{1/\lambda,1/\lambda^2,1/\sqrt 2\}$ and $\{\ln 3,\ln 4,\ln 5\}$, with $\lambda^3=1+\lambda$. Both sets gave equivalent results, within the numerical accuracy. Data given in the text is for the first set of frequencies.
\bibitem{note3} Eq.\ \eqref{nusc} was derived in Ref.\ \cite{Gar08} for a time-reversal symmetry operator that squares to $+1$ (orthogonal symmetry), while we have ${\cal T}^{2}=-1$ (symplectic symmetry). Apparently the critical exponent is not sensitive to this difference.
\end{thebibliography}
\end{document}